\documentclass[%
 reprint,
 amsmath,amssymb,
 aps,
]{revtex4-1}

\newcommand{\mc}{\mathcal}
\usepackage{color}

\usepackage{hyperref}
\usepackage{graphicx}
\usepackage{dcolumn}
\usepackage{bm}
\usepackage{color}

\begin{document}
\bibliographystyle{naturemag}

\preprint{APS/123-QED}

\title{Emerging Frontiers of Neuroengineering: A Network Science of Brain Connectivity}

\author{Danielle S. Bassett$^{1,2}$}
\author{Ankit N. Khambhati$^{1}$}
\author{Scott T. Grafton$^{3,4}$}
\affiliation{
 $^1$Department of Bioengineering, University of Pennsylvania, Philadelphia, PA, 19104
}
\affiliation{
$^2$Department of Electrical and Systems Engineering, University of Pennsylvania, Philadelphia, PA, 19104
}
\affiliation{
 $^3$ UCSB Brain Imaging Center and Department of Psychological and Brain Sciences, University of California, Santa Barbara, CA 93106 USA
}
\affiliation{
$^4$ Institute for Collaborative Biotechnologies, University of California, Santa Barbara, CA 93106 USA
}

\date{\today}
\begin{abstract}
Neuroengineering is faced with unique challenges in repairing or replacing complex neural systems that are composed of many interacting parts. These interactions form intricate patterns over large spatiotemporal scales, and produce emergent behaviors that are difficult to predict from individual elements. Network science provides a particularly appropriate framework in which to study and intervene in such systems, by treating neural elements (cells, volumes) as nodes in a graph and neural interactions (synapses, white matter tracts) as edges in that graph. Here, we review the emerging discipline of \emph{network neuroscience}, which uses and develops tools from graph theory to better understand and manipulate neural systems, from micro- to macroscales. We present examples of how human brain imaging data is being modeled with network analysis and underscore potential pitfalls. We then highlight current computational and theoretical frontiers, and emphasize their utility in informing diagnosis and monitoring, brain-machine interfaces, and brain stimulation.  A flexible and rapidly evolving enterprise, network neuroscience provides a set of powerful approaches and fundamental insights critical to the neuroengineer’s toolkit.
\end{abstract}

\maketitle

Could we graft new connections into the brain, to give someone back the abilities they had pre-injury \cite{chen2016neural}? Could we decode the thoughts of someone who is caged inside their own body \cite{haynes2006decoding,christophel2012decoding}? Could we develop adaptive brain-computer interfaces that evolve and adapt to remain effective for a child whose brain is continuously developing \cite{putze2014adaptive,krusienski2011critical}? Answering these and many other seemingly over-ambitious questions is the fundamental aim of neuroengineering \cite{dilorenzo2007neuroengineering}, a relatively new domain of biomedical engineering that develops and uses computational and empirical techniques to understand and modulate the properties of neural systems. Particularly exciting frontiers of neuroengineering include neuroimaging, neural interfaces, neural prosthetics and robotics, and more general techniques for regeneration, enhancement, and refinement of neural systems \cite{johnson2013neuromodulation}.

In the era of big data, neural systems are no exception to the rule of ever-increasing petabytes streaming in to servers around the world  \cite{glaser2016development}. However, in many other arenas, the amount of data being gathered has not posed an insurmountable obstacle. What is the fundamental difference that causes neuroscientific data to be so challenging? Is it a lack of a mechanistic understanding of
how the brain works \cite{valiant2014what,craver2005beyond}? Or an inability to physically construct the hardware required to liaise with neural systems for effective interventions \cite{krusienski2011critical}? We argue that fundamental to both of these problems is the challenge of dealing with complex relational data \cite{bullmore2009generic}. In developing a data science to meet these rising demands \cite{stevenson2011how}, we must acknowledge that these data are far from independent: instead, data from neural systems are inherently relational data \cite{bassett2016network}.

\begin{figure}[h]
\begin{center}
\centerline{\includegraphics[width=0.45\textwidth]{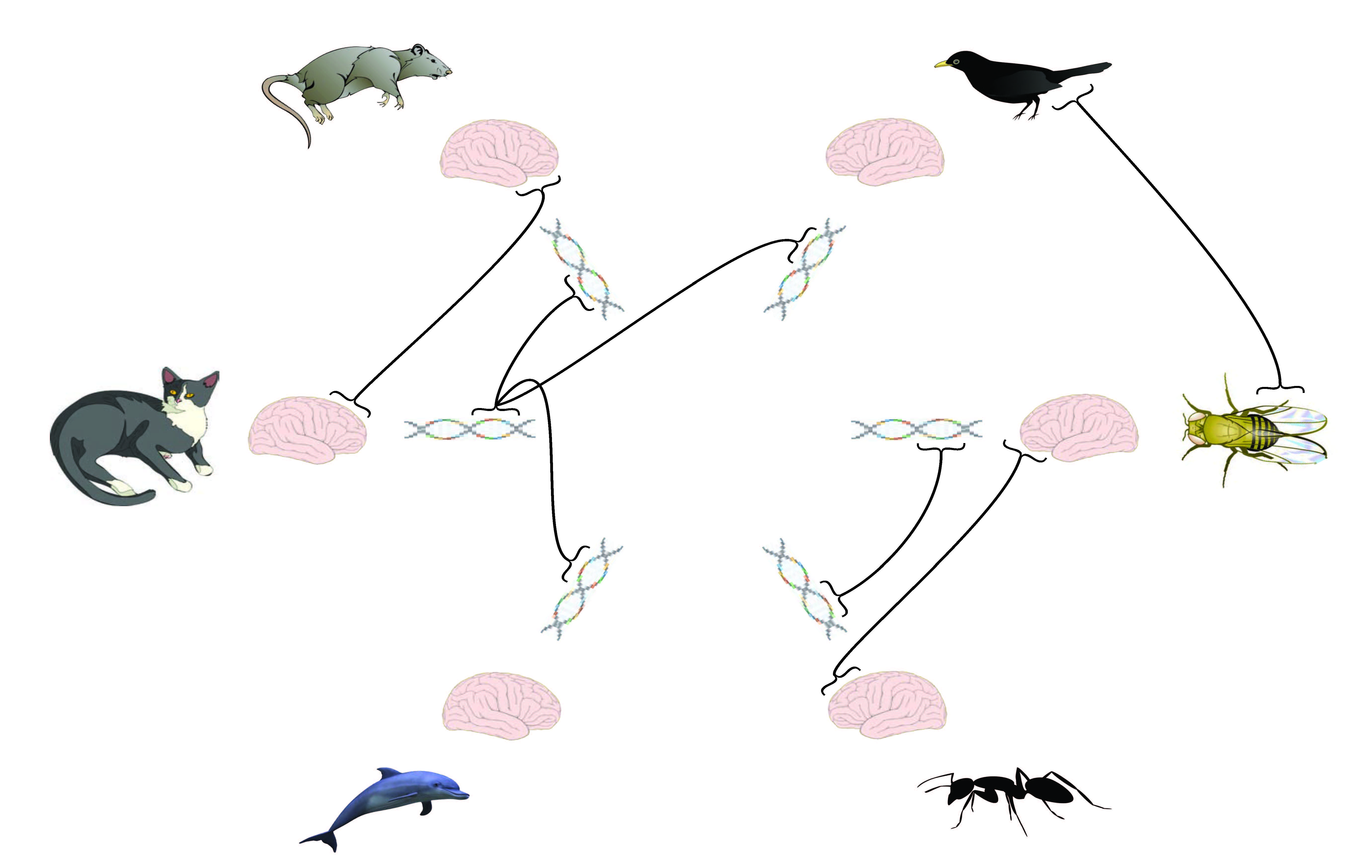}}
\caption{\textbf{Relational data in biological systems.} Repeating genotypic and phenotypic patterns emerge frequently in the study of biological systems. These biological patterns are expressed across multiple scales of granularity. Illustrated here are three different scales of biological elements (behavioral, structural, genetic) in different animal species, with lines representing conceptual relationships between elements. At the macro-scale, we observe behavioral similarities across different species, such as the ability to fly in birds and fruit flies. However, a closer lens on the neurological substrate of this behavior may tell a different story: that meso-scale structural brain architecture differs significantly between birds and fruit flies, and is more similar between insects (e.g. fruit flies and ants) and between mammals (e.g. mice and cats). Despite differences in structural brain architecture, we might find that animals of different species share commonalities in genetic code that manifest similarly in physical attributes. While differences in each element yield unique qualities to each individual animal species, examining relational data can provide a more comprehensive view on the functional role of each element ecologically.} \label{fig1_multiscale}
\end{center}
\end{figure}

Relational data can be defined as any data that codifies relationships between elements  \cite{long2010relational}. The nervous system is composed of units across many spatial scales (genes, neurons, columns, areas) that are related to one another in many different ways (anatomical connections, functional relationships, material similarities) \cite{conaco2012functionalization,zhang2016stretch} (Fig.~\ref{fig1_multiscale}).  These relationships form intricate patterns -- of synaptic connections, gene co-expression, connectome fingerprints -- that may differ across species \cite{van2016comparative,shih2015connectomics}, or across cohorts within a single species (e.g., in health \emph{versus} disease) \cite{bassett2009human,fornito2015connectomics}.  From these patterns stem the very complicated phenomena of development, behavior, and cognition \cite{medaglia2015cognitive,misic2016from}.

Biological patterns like these are particularly difficult to study for several reasons. First, the governing principles of pattern formation are often difficult to infer \cite{vertes2012simple,vertes2014generative,betzel2016generative}, and thus mechanistic insights are difficult to come by. Second, it is difficult to simplify patterns using coarse-graining or other dimensionality reduction approaches while still maintaining the richness of the neurophysiologically relevant information \cite{craver2005beyond}. While retaining necessary information while simplifying the patterns is difficult, so is studying each element in the pattern: with thousands and sometimes millions of elements, the set of interactions between them -- particularly if they evolve in time -- quickly becomes enormous and complicated.  Indeed, as many fields have now come to realize, the intricacies of relational data call for a new conceptual and mathematical framework \cite{pilosof2015multilayer,proulx2005network}.

\section*{The Peculiar Appropriateness of Network Science}

Network science is an emerging interdisciplinary field that combines theories from statistical mechanics, computational techniques from computer science, statistics, applied mathematics, and visualization approaches to probe, perturb, and predict the behavior of complex systems in technology, biology, and sociology \cite{newman2010networks}. While historically developed to understand social interactions or friendship webs similar to those supported (or elicited) by Facebook or Twitter, network science is a peculiarly appropriate framework in which to tackle the challenges of neural data sciences to better engineer artificial and natural neural systems.

In particular, rather than reducing complex relational data to a list of independent parts, network science provides tools to explicitly characterize the pattern of interactions between neural elements \cite{rubinov2010complex}. In addition to these descriptive tools, it provides benchmark graphical models for statistical comparison and inference \cite{pavlovic2014stochastic,simpson2011exponential,lindquist2014evaluating}, mathematical models to quantify and predict the flow of information \cite{misic2015cooperative} or communication \cite{misic2014communication} through neural circuits, and predictive tools to forecast how networks might change in response to injury \cite{patel2014single} or therapeutic interventions \cite{gratton2013effect}.

How does one go about translating neural systems into the language of network science \cite{kaiser2011tutorial,fornito2013graph,vandiessen2015opportunities,reijneveld2007application,bullmore2009complex,bassett2009human,bassett2006small,bullmore2011brain}? The first critical step is to determine which constituent elements are the fundamental unit of interest that is measurable in the particular experiment under consideration \cite{butts2009revisiting}. These elements -- which might be single neurons, neuronal ensembles, genes, or large-scale brain areas -- will be treated as nodes or vertices in the network \cite{zalesky2010whole}. Then, one must define the connections, interactions, or relationships of interest between network nodes. These links -- which might be white matter tracts between large-scale brain areas, chemical or electrical synapses between neurons, or co-expression patterns among genes -- will be treated as network edges. Once nodes and edges have been defined, the network itself -- the pattern of edges linking nodes -- can be studied from the point of view of a graph in mathematics \cite{bollobas1985random,bollobas1979graph}.

\section*{The Mathematics of Network Science in Neural Systems}

\begin{figure}[h]
\begin{center}
\centerline{\includegraphics[width=0.45\textwidth]{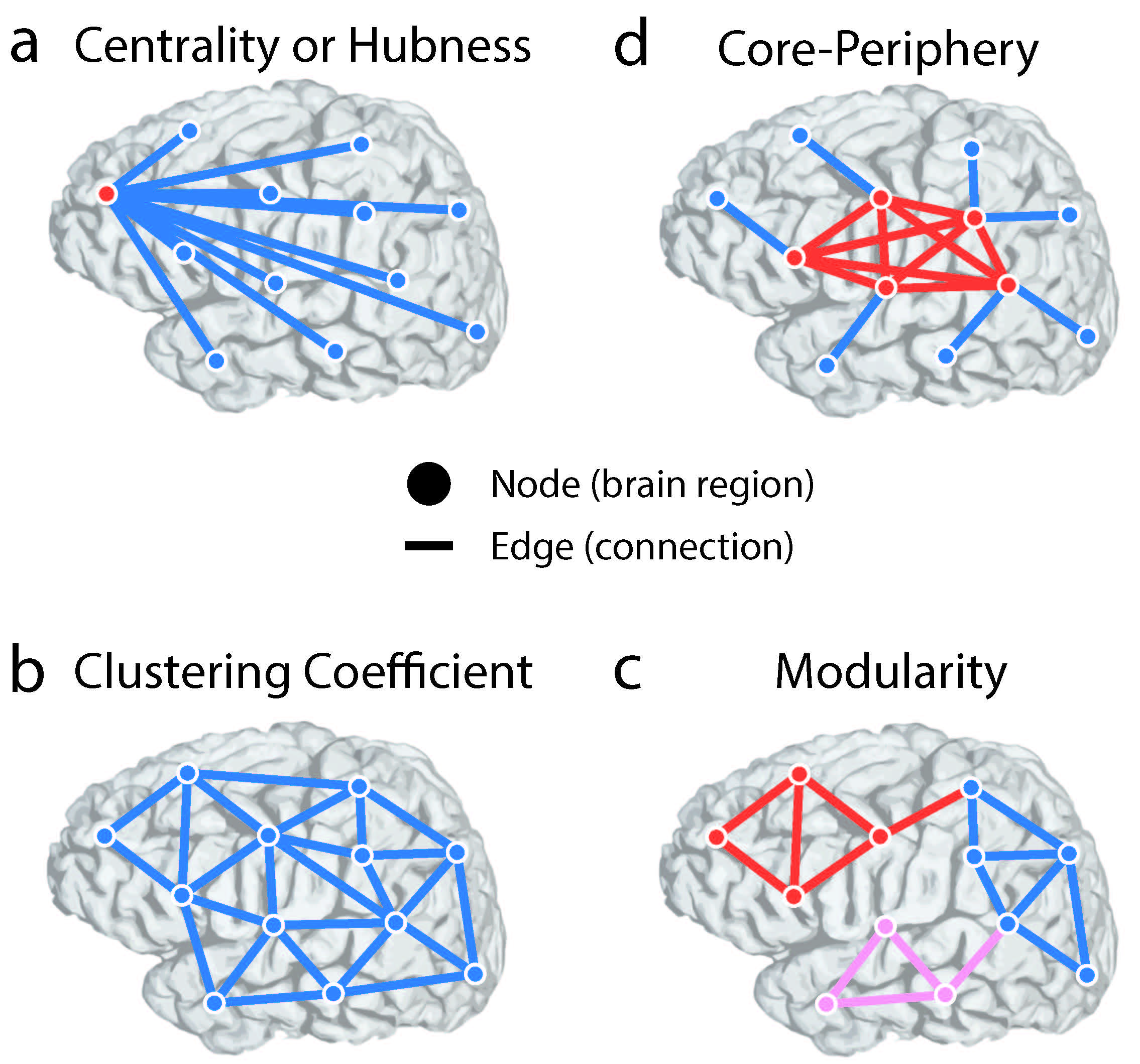}}
\caption{\textbf{Multi-scale topology in brain networks.} Brain networks express fundamental organizing principles across multiple spatial scales. Brain networks are modeled as a collection of nodes -- representing regions of interest with presumably coherent functional responsibilities -- and edges -- structural connections or functional interactions between brain regions. (\textbf{A}) Node centrality describes the importance of individual nodes in terms of their connectivity relative to other nodes in the network. Nodes with more connections or stronger edges tend to be hubs (red), while nodes with less connections tend to be isolated (blue). (\textbf{B}) Clustering coefficient, a measure of connectivity between the neighbors of a node, is another local measure of network topology. Unlike network topologies with strong hubness qualities, as in \textbf{A}, networks with strong clustering coefficient demonstrate a high density of triangles that is believed to facilitate local information processing. (\textbf{C}) Modularity is a meso-scale topological property that captures communities of nodes that are tightly connected to one another and weakly connected to nodes in other communities. Modular organization underlies a rich functional specialization within individual communities. Here, nodes of different communities are colored red, blue, or pink. (\textbf{D}) Networks with core-periphery structure exhibit a set of tightly-connected nodes (core; red) sparsely connected to a set of isolated nodes (periphery; blue). This organization is in stark contrast to the modular organization in \textbf{C}. The core-periphery architecture is characteristic in networks that integrate information from isolated regions in a central area.} \label{fig2_network_structures}
\end{center}
\end{figure}

In the field of mathematics, a graph $\mc G = (\mc V, \mc E)$ is composed of a node set $\mc V$ and an edge set $\mc E$ \cite{bollobas1979graph,bollobas1985random}. We store this information in an adjacency matrix $\mathbf{A}$, whose elements indicate the strength of edges between nodes. The representation of data in a graph enables the investigator to characterize the patterns of connectivity locally surrounding a single node or globally taking into account all edges. In addition to local and global structure, tools are available to probe so-called \emph{mesoscale} structure in the graph, which can be defined as structure that is present at intermediate length scales in the system (Fig.~\ref{fig2_network_structures}).

To give the reader some simple intuitions, we briefly describe examples of local, meso-scale, and global statistics that can be computed from graphs of neural systems. First, a common \emph{local} statistic that has proven particularly effective in characterizing neural systems is the \emph{clustering coefficient} of a node, which can be defined as the fraction of a node's neighbors that are also connected to one another \cite{watts1998collective}. In essence, this statistic is sensitive to the density of triangles in the graph, and is thought to play a non-trivial role in local information processing in neural systems \cite{kitzbichler2011cognitive} (although for caveats see \cite{rubinov2011emerging,bassett2015cognitive}). A common \emph{global} statistic that has proven useful in characterizing neural systems is the  \emph{characteristic path length}, which is defined as the average shortest path between all possible node pairs \cite{newman2010networks}. This statistic is sensitive to long-distance connections that provide short cuts from one side of the network to another, and is thought to play a role in the swift transmission of information across the system \cite{bullmore2011brain}. Interestingly, early work demonstrated that humans displaying brain wiring patterns with shorter characteristic path length also had higher IQ than those with longer characteristic path length \cite{li2009brain}, suggesting the sensitivity of network statistics to architectures that support healthy human cognitive function. However, it is worth noting that short characteristic path lengths do not appear to be the full story \cite{bassett2006small,bassett2016small}, and measures of segregated information processing also play an important role in brain function \cite{sporns2016modular}.

In addition to local and global structure, meso-scale organization provides a window into the properties of groups of nodes. Two common mesoscale structures are modularity and core-periphery structure. A network with modular structure is one that contains groups of nodes also known as modules; the nodes in a module are more densely connected to other nodes in the same module than to nodes in other modules \cite{porter2009communities,fortunato2010community}. This modular architecture is thought to support specialization of function, each module performing a different role in support of neurophysiological processes from synchronization to cognition \cite{sporns2016modular}. In contrast, a network with core-periphery structure contains a set of core nodes that are densely interconnected with \emph{all} other nodes in the network, and a set of periphery nodes that are sparsely interconnected with all other nodes in the network \cite{everett1999peripheries,borgatti1999models}.  This organization is thought to support the integration of information across neuronal assemblies, neural circuits, or large-scale functional modules \cite{bassett2013task}, in each of which the top-down web os often referred to as a \emph{rich club} \cite{van2011rich}.

The multiscale nature of these network tools are particularly useful for neural systems, which are thought to perform inherently different computations at different levels of the network hierarchy \cite{bassett2013multiscale}. For example, information is thought to be processed in local cortical areas before being passed across modules along rich-club edges (in a so-called \emph{small-world} \cite{bassett2006small,bassett2016small}), allowing integrative computations and coherent behavioral responses \cite{bassett2010efficient,bassett2011conserved}. Understanding this multiscale architecture and its functional role in neural system dynamics is critical for developing effective interventions that capitalize on existing structure and dynamic properties rather than fighting against them.

\section*{How do we build brain networks?}

Using the mathematical tools of network science to understand neural data requires one to explicitly build network models. How does one go about doing so? This topic fully warrants a review of its own: describing methods to build brain networks from spiking data \cite{muldoon2013spatially}, calcium transients and microelectrode arrays \cite{bettencourt2007functional}, mesoscale tract tracing \cite{bassett2016small}, genetic expression \cite{fulcher2016transcriptional}, and large-scale neuroimaging \cite{bullmore2011brain} and across species from cat \cite{hilgetag2004clustered} and macaque \cite{markov2013cortical}, to \emph{C. elegans} \cite{bassett2010efficient}, mouse \cite{rubinov2015wiring}, rat \cite{heuvel2016topological}, drosophila \cite{kaiser2015neuroanatomy,shih2015connectomics}, and human \cite{hagmann2008mapping} (to offer a sparse list!). Because we cannot do justice to the full richness of this question here, we focus our presentation on human brain imaging data, which has historically provided the largest source of data for testing the utility of network science to characterize complex neural systems. Thousands of healthy subjects and patient populations have been scanned, primarily by magnetic resonance imaging to identify both structural and functional properties of the nervous system. Given the dominant role of imaging, here we familiarize the reader with the strategies used to transform brain scans into data structures that are amenable to network analysis. However, we emphasize that the network tools we describe are fully translatable to other neural systems, and are commonly applied in EEG \cite{deng2015brain,toppi2015graph,zhang2013prediction}, MEG \cite{bassett2009cognitive}, ECOG \cite{khambhati2015dynamic,khambhati2016virtual}, and fNIRS \cite{niu2012revealing,zhang2016mapping} as well (for a more thorough review of these application areas, see \cite{bassett2016network}).

\begin{figure*}[t]
\begin{center}
\centerline{\includegraphics[width=0.75\textwidth]{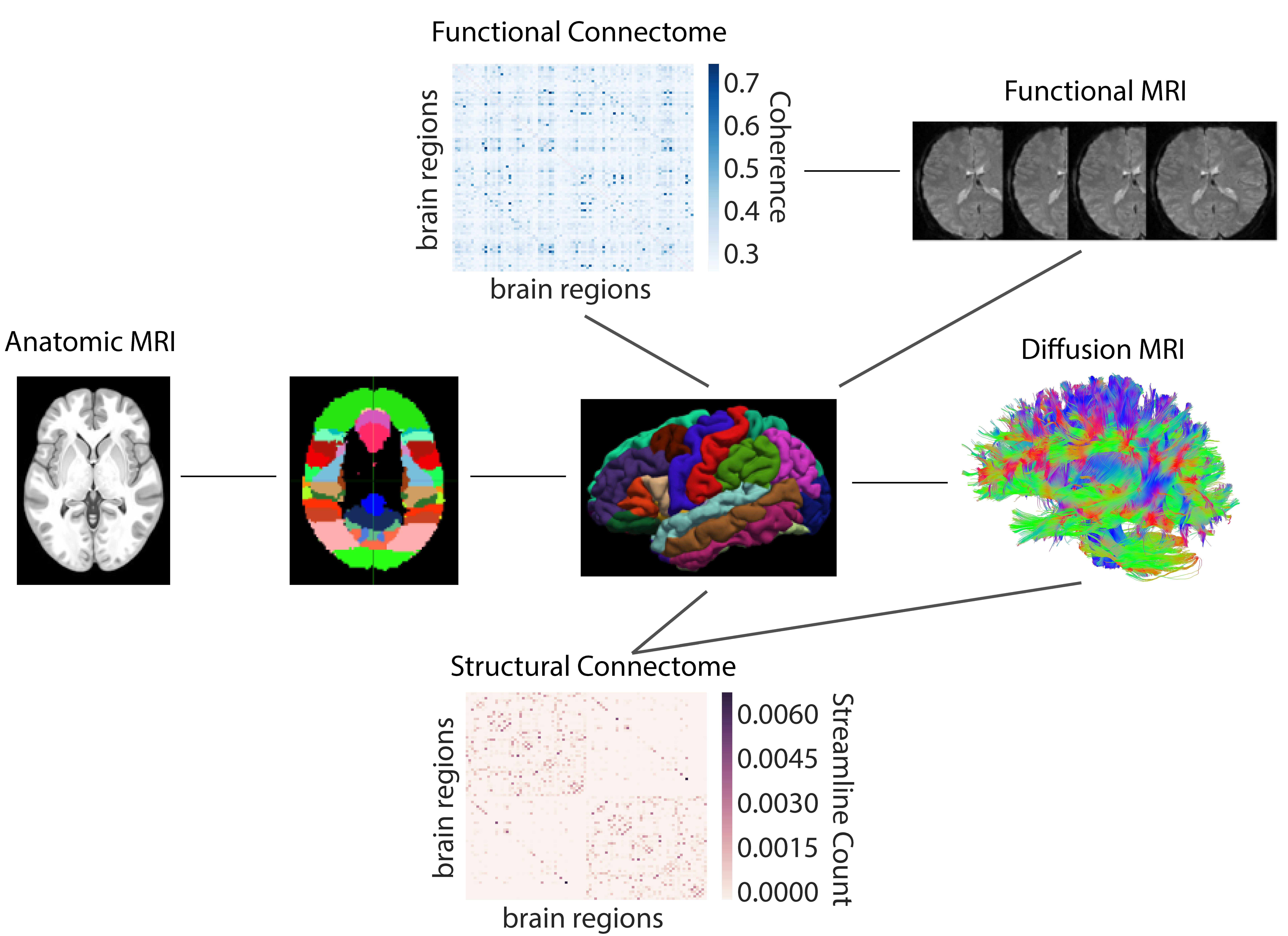}}
\caption{\textbf{Constructing Connectomes from MRI Data}. To generate human connectomes with magnetic resonance imaging, an anatomic scan delineating gray matter is partitioned into a set of nodes. This is combined with either diffusion scans of white matter structural connections or time series of brain activity measured by functional MRI, resulting in a weighted connectivity matrix.} \label{fig_stg}
\end{center}
\end{figure*}

The most common measurements of human brain connectivity, whether they are functional or structural networks, rely on scans obtained by magnetic resonance imaging (MRI). There are three basic types of brain scans that are typically used in network construction. The first is an ``anatomic'' scan. This is a T1 weighted, high resolution ($<$1mm isotropic) sequence that can distinguish gray matter from underlying white matter. Many software tools are available for segmenting these two types of tissues and for partitioning the gray matter into a set of local regions that form the network nodes, as shown in Fig.~\ref{fig_stg}. There are many atlases available for partitioning the gray matter, varying from $\sim$50-1000 separate regions. The second type of scan is a ``functional'' MRI. This is a series of T2* weighted scans acquired at sampling rates as fast as 2.21 Hz (although more typically at 0.5 Hz) and at a lower spatial resolution ($\sim$3mm non-isotropic). This tissue contrast is sensitive to changes in blood oxygen level dependent (BOLD) signals, which vary as a function of cortical activity (whether neuronal or synaptic). These scans can be acquired with the subject at rest, the so-called ``resting state'' MRI or while performing a particular task, the so-called ``task based'' fMRI  \cite{RN13458,RN15502}. The time series of brain activity, averaged across all voxels in each local region can then be extracted. To create an adjacency matrix reflecting functional connectivity, pairs of time series are related by correlation, partial correlation, wavelet filtered correlation or coherence within a particular frequency band.

The third method of imaging is based on diffusion imaging. This involves the acquisition of a set of scans, each of which is sensitive to the magnitude of water diffusion in a particular direction in 3-dimensions. The set of oriented diffusion scans are then combined to estimate voxel-wise distributions of water diffusion \cite{RN9963}. The brain is then seeded uniformly at subvoxel resolution and a probable path of diffusion through the full volume is calculated, resulting in a virtual tract of diffusion, referred to as a ``streamline''. These streamlines are virtual estimates of possible water diffusion that can correspond to true white matter fascicles or tracts. The white matter fascicles and tracts are thought to be the primary means for information sharing between distinct gray matter regions, analogous to wiring that connects distinct computer modules \cite{hagmann2008mapping,bassett2011conserved}. To create an adjacency matrix reflecting this structural connectivity, the strength of connectivity between pairs of regions can be estimated by the number of streamlines, the density of streamlines or the number of streamlines normalized by the length they traverse. The set of all real connections in the brain is referred to as the human connectome. While it is not yet possible to characterize this full set of connections, it can be approximated by the streamlines reconstructed with diffusion imagng. This lower dimensional connectome can then be characterized with the tools of network science. Given that it is an approximation, it is valuable to test the robustness of any particular network property across a range of atlases or spatial resolutions.

\section*{What do brain networks offer neuroengineering?}

After building networks from imaging data, one can then use these networks to address pressing questions in neuroengineering. While we cannot exhaustively cover all possible uses of these tools currently in the literature, here we highlight their utility in neural mapping and connectivity estimation, diagnosis and monitoring, and rehabilitation and treatment.

\subsection*{Diagnosis and monitoring}

Accurately diagnosing disorders of the human connectome and monitoring their progression are particularly critical applications of network-based tools to neural systems. Diseases thought to be accompanied by connectome abnormalities or alterations include schizophrenia \cite{bassett2008hierarchical,bassett2009cognitive,lynall2010functional}, autism \cite{nomi2015developmental,menon2011large}, epilepsy \cite{burns2014network,khambhati2016virtual}, and Alzheimer's disease \cite{he2009neuronal,poza2013characterization,tijms2013alzheimers}, among others \cite{stam2014modern,bassett2009human}. The pattern of alterations in a given condition can be described in the form of a graph, as can the patterns that are similar or different between a pair of conditions. In some cases, these network changes occur early in a disease, offering potential as diagnostic biomarkers \cite{zhu2016changes}. Bolstering this possibility, several studies have demonstrated that by incorporating network statistics as features in machine learning algorithms, it is possible to classify groups of individuals with and without a particular condition, from aging \cite{petti2013aged} to major depression \cite{sacchet2014elucidating}. While diagnosis and classification are binary decisions, one can also continuously monitor brain networks within a single individual \cite{toppi2014investigating} either during drastic changes such as those accompanying disease progression, or during minor changes in mental state such as those induced by driving fatigue \cite{zhao2016reorganization}.

\subsection*{Rehabilitation and treatment}

The sensitivity of network measures to brain state offers the generalizable potential for graph statistics to be used as indicators of the efficacy of rehabilitation and treatment. Initial studies support this potential efficacy by demonstrating appreciable changes in network organization induced by memory rehabilitation treatment (a broad intervention useful across multiple clinical conditions \cite{toppi2014time}), seizure therapy (an intervention for severe depression \cite{deng2015brain}), and motor imagery (a frequent intervention for stroke \cite{ge2015motor}). Common techniques to affect these interventions include neurofeedback where humans learn to control the activity or connectivity in certain parts of their brain to enhance mental function \cite{stoeckel2014optimizing,banca2015visual}. Notably, graph statistics of functional network architecture have proven sensitive to cognitive workload during these interventions, offering task-independent markers for monitoring and matching participants ability and task difficulty during neurofeedback training \cite{fels2015predicting}. Neurofeedback approaches often utilize exquisitely calibrated brain-computer interfaces \cite{andersen2014toward}, systems that can also dual as neural prosthetics. When applying these techniques to clinical populations, a pressing question arises due to limited resources: Who will benefit most? Can we choose the intervention that best fits a given individual? Interestingly, emerging data suggest that organizational characteristics of a person's functional network architecture as measured by EEG can be used to predict who will be receptive to motor imagery treatment \cite{zhang2015efficient,zhang2013prediction}. These initial studies underscore the potential of network representations of neural data to provide sensitive and specific markers of the receptiveness of neural circuits to induced network structure change \cite{sporns2013human}.

\subsection*{Neural mapping and neural connectivity estimation}

While neuroengineering is often thought of as a field of clinical translation, basic science plays a fundamental role in giving the investigator the knowledge and understanding necessary to intervene in a way that benefits the system. A particularly exciting current frontier in neuroengineering lies in mapping neural systems using a variety of imaging techniques \cite{christopoulos2012network}, and in estimating the connectivity between neural elements using sophisticated statistical algorithms \cite{kafashan2015optimal}. In these contexts, network science offers explicit tools to characterize the maps, and to use empirical estimates of connectivity to inform the design of new networks. Indeed, the concept of network design is a relatively new one in biological systems. When applied to the neural domain, network design includes the building of computational models of neural dynamics, as well as physical models \cite{kanagasabapathi2012functional} via micropatterning, microfabricated multielectrode arrays, and low-density neuronal culture techniques \cite{chang2006neuronal}. Together, these algorithmic and empirical approaches provide exciting avenues to map the neural connectome across scales and species \cite{bassett2016network}, and to better understand the dynamics that produce cognition and behavior \cite{medaglia2015cognitive}.

\section*{Constructing and Using Brain Networks for Neuroengineering}

This brief survey of the literature demonstrates that brain networks offer exciting capabilities in addressing pressing questions in neuroengineering. In this section, describe important considerations in constructing and using brain networks in the context of human imaging. While we focus on human neuroimaging, these (or similar) considerations are likely to be important in the collection and analysis of other types of data (multi-unit recording, optical imaging) as they become available for network analysis.

\subsection*{Image Acquisition}

The rapid growth of imaging-based network science has been accompanied by a parallel recognition that functional and structural MRI data can be corrupted by a broad array of technical, physiologic and anatomic factors that, if not handled properly, lead to major errors in network modeling. The good news is that MRI is a mature technology and it is rare for data to be corrupted by artifacts secondary to unreliable hardware or poor pulse sequence designs. The bad news is that brain imaging is commonly corrupted by more subtle physical-anatomic properties that can be difficult to surmount with conventional hardware and pulse sequences \cite{RN15485}. Both diffusion weighted imaging (used for structural connectomics) and T2* weighted imaging (to detect changes of BOLD signals in functional connectomics) are highly sensitive to susceptibility artifacts. The most troublesome cause is an air-tissue interface that leads to very localized non-linear image distortion and signal irregularity. For example, the medial and inferior temporal and orbital frontal cortex of the brain are adjacent to air filled petrous and ethmoid sinuses. The resulting artifacts lead to missing streamlines projecting into these areas or unreliable estimates of functional activity within the distorted gray matter regions. These distortions are difficult to correct \emph{post hoc}. The degree of signal dropout and missing data varies enormously between individual subjects. Thus, network analyses that are aggregated over a population need to carefully evaluate the influence of missing data on the underlying connectivity matrices. A second challenge in brain MRI, particularly for diffusion imaging is the effect of eddy currents. Eddy currents are loops of electrical current induced within the brain tissue by the changing magnetic field required to generate images. This causes spatial distortion within each image slice and can be particularly impactful in diffusion imaging. A third challenge, also leading to geometric distortion, arises from magnetic field inhomogeneity and phase encoding errors. There are numerous software tools available for correcting both types of distortion \emph{post hoc}.

\subsection*{Pitfalls unique to functional imaging networks}

Ideally, all of the functional connections, whether for a resting state \cite{RN15491,RN15490,RN15489} or task-based network would be determined by patterned brain activity reflecting inherent cognitive processes. However, there are numerous other sources of noise that can contribute to spurious increases of functional connectivity \cite{RN15470,RN15457,RN15464}. One of the most important influences on functional connectivity is variations in amplitude or rate in the cardiac and respiration cycles. Respiration rate ($\sim$0.2 Hz) and depth of breathing can clearly influence local BOLD signal \cite{RN13210}. Cardiac rate ($\sim$1 Hz) also influences BOLD signal \cite{RN15492}. These effects are regionally complex, with respiration effects more apparent near the ventricles, and cardiac effects more apparent near the largest arteries. Higher order effects of the cardiac and respiratory cycle may also be present in the tissue beyond a simple linear projection of the pulse and bellows signals. For example, the chest wall expansion will influence global magnetic field inhomogeneity, while CSF pulsatility (via the cardiac pulse) may be periodic with the chest expansion. Thus, the influence of both on functional connectivity analyses will be dependent on an individual subject's physiologic state and unique anatomy.

There are many retrospective strategies for removing the effects of cardiac and respiratory cycle variation on the BOLD time series from each voxel. If heart rate and respiration have been independently measured, then software tools such as ``RETROICOR'' can be used \cite{RN15493}. It uses a Fourier expansion of the non-brain physiologic signals with 8-20 regressors. While this method works well for both linear and higher order artifacts, there is a trade-off in that increasing the number of regressors during RETROICOR correction will remove a greater amount of relevant brain signal \cite{RN15461}.

For many experimental situations, independent measures of heart rate and respiration are not available and methods besides RETROICOR are needed. Here we mention techniques based on independent component analysis (ICA) of the rsMRI data. ICA decomposes the functional time series for all voxels into patterns of activity consisting of a set of spatial maps, each of which has a corresponding time series that when added linearly, sum to the original voxel-wise time series. A set of ICA components will represent both brain activity and “noise” components. Ideally, these sources of brain and non-brain activity are independent. If so, then these latter noise components can be removed and a new noise free times series can be reconstructed. The challenge then, is to find an unbiased, efficient method for identifying those components reflecting noise. Manual classification of ICA components is very difficult, and requires expert knowledge. One semi-automated ICA-based X-noiseifier called ``FIX'' \cite{RN15421} uses a machine learning approach to aid with this process. For each ICA component a large number of distinct spatial and temporal features are generated, each describing the proportion of temporal fluctuations observed at high frequencies. These features are fed into a multi-level classifier. After training by hand-classification across a sufficient number of datasets, the classifier can then be used with new datasets.

An alternative approach is to estimate pulse and respiratory variability for a subject directly from an independent set of fMRI data, utilizing temporal independent component analysis \cite{RN15460}. The method assumes that non-brain physiologic noise is spatially stationary. For example, noise associated with the carotid arteries will be in the same location across different rsMRI scans from the same subject. Once the underlying and independently derived spatial weighting matrix is identified by ICA in one dataset, it can be applied to a separate rsMRI time series from the same subject to produce the temporal pattern of noise. The resulting cardiac and respiratory estimators can then be used with RETROICOR or similar correction methods. While this method works well, it requires an independent sample of functional data.

It has been assumed that global BOLD activity, measured over the whole brain, will remain constant across a time series. Any fluctuation would be due to instrumentation issues or non-brain physiologic effects. However, recent studies have examined the effect of removing the global mean signal from the time course on subsequent connectivity analyses. Interestingly, multiple studies show that a significant portion of the global mean signal is in fact related to the average signal within particular resting state brain networks \cite{RN15494,RN15497} and that removing the global signal can result in spurious negative correlations \cite{RN13032} and reduces reproducibility of many network metrics \cite{RN15471}. Despite these disadvantages, global signal regression can be helpful in developmental and clinical cohorts to correct for motion-related artifact \cite{RN15500,ciric2016benchmarking}.

Indeed, it is almost impossible for a person to remain motionless in an MRI scanner. Breathing, swallowing and volitional movements can create motion that propagates to the head. A brain placed in the MRI field will become magnetized over $\sim$6 seconds. If the brain moves, then it will no longer be magnetized in the same direction and there can be a massive increase in the signal until the brain has remagnetized to the new magnetic orientation. To account for the effects of this motion-induced noise, a variety of retrospective methods have been proposed. Most assume that the change in signal intensity will be global, occurring within a single time sample of the rsMRI time series. One of the most common methods is to use linear transformations to fit each time sample to one time point. The resulting transformation weights (translations and rotations) can be used to adjust global signal intensity or be included in a regression model as a covariate of non-interest \cite{RN15499,RN15498}. However, the ``filtering'' of time series data with motion parameters is problematic because they do not model continuous motion directly. Rather, they capture net displacement at the temporal resolution of the sampling frequency ($\sim$0.5 Hz). If the head is displaced rapidly, and returns to the same position within a single sampling period, then there is no net displacement, but a large signal spike in the data. This will profoundly alter the strength of connectivity between areas with a common motion-induced signal change. This type of signal change has been described as the ``predominant effect of motion'' in a sample ranging from 8 to 23 years old \cite{RN15500}. In recognition of potential artifacts from rapid motion (or RF spikes), software has been developed to address them. Rather than using the realignment information, these methods search for global spikes in signal intensity. There is one final challenge with head motion artifacts. Within each volumetric acquisition, a stack of slices are acquired sequentially, typically by first sampling the odd slices and then the even slices. It is not uncommon for a brief head movement to demagnetize a subset of slices, causing artifacts in every other slice. This is not detected in the transformation matrix and may not alter global signal intensity. New tools are emerging to detect unexpected signal spikes within single slices \cite{RN15472}.  For a useful study benchmarking confound regression strategies for the control of motion artifact in studies of functional connectivity, see \cite{ciric2016benchmarking}.

\subsection*{Pitfalls unique to structural imaging networks}

There are many sampling schemes for acquiring a set of oriented diffusion scans. These include diffusion tensor imaging (DTI), which sample an object at a uniformly spaced set of angles and at a constant magnetic gradient strength. When the gradient strength or number of directions are increased the angular resolution improves (Q-ball and high angular resolution diffusion imaging ``HARDI'') but with the tradeoff of reduced signal to noise in the scans.  Multiple shells of gradient strengths can be applied (multishell diffusion imaging) or a uniform distribution across gradient strength and direction can be applied (diffusion spectrum imaging) \cite{RN15488}. Critically, each of these methods requires a different mathematical technique for converting diffusion images to probabilistic estimates of local water diffusion, resulting in varying success at modeling the connectivity in different brain areas where there can be water diffusion in multiple directions (the crossing fiber problem). Methods using lower angular resolution such as DTI consistently underestimate the number of possible streamlines by an order of magnitude compared to multishell and DSI methods. Missing connections can also arise because an insufficient number of seeds are introduced to generate the underlying streamline set. Whatever the cause, allowing for missing data can significantly alter graph metrics \cite{RN13458}. On the other hand, commonly used algorithms for generating streamlines can create noisy, anatomically implausible connections that must be removed by length and/or angle thresholds. Most current algorithms for generating streamline connections suffer from a length bias \cite{RN15501,RN15503}. The shorter a connection, the easier it is to be reconstructed. Thus, a structural network will be more likely to represent short connections than long. Similarly, the odds are more likely that a streamline will be reconstructed if it is in a thick white matter fascicle with many fibers oriented in a common direction than if it is in a thin fascicle. To address the length bias, some authors normalize the streamline count between two gray matter regions by the physical size of those same regions. Clearly, standardization in these acquisition, reconstruction and counting procedures is essential for reproducibility and generalizability.

\section*{Frontiers in Computational Science and Systems Engineering}

With these empirical considerations in mind, it is nevertheless clear that brain imaging has provided a fertile test bed for developing and testing novel tools from network science. Yet, it is likewise clear that this is only a first wave of innovation. Indeed, network neuroscience offers to the field of neuroengineering two distinct sets of frontiers: one in the development of computational and systems engineering approaches, and the other in translating current and future advances directly to clinical populations. In this section, we briefly review new directions in algorithmic development, computational architectures, signal processing techniques, and statistics that support the extraction, representation, and characterization of meaningful relational patterns in neural data. We also discuss the nascent application of control theory to these networks, and highlight their potential utility in guiding clinical interventions.

\subsection*{Dynamic and multilayer networks}

A commonly faced challenge in applying network analyses to neural data is that the processes we often wish to understand are inherently \emph{dynamic} processes \cite{hutchison2013dynamic,calhoun2014chronnectome,kopell2014beyond}. Rehabilitation, response to treatment, monitoring disease progression, and tracking BCI learning are all evolving processes that occur over a range of time scales. Yet, networks in their simplest forms are static: a fixed set of network nodes are connected by a single estimate of connectivity. In extending these static descriptions to incorporate time, the applied mathematics community has defined so-called \emph{multilayer networks} \cite{kivela2014multilayer}. Colloquially, a multilayer network is a network that contains different layers, and in which the edges in a given layer represent a different type of relationship than the edges in another layer. Perhaps the simplest type of multilayer network is a \emph{temporal network}, where each layer is a time window and the edges within that layer represent relationships that are true in that time window \cite{holme2012temporal}. By tying each layer to the next using identity links (edges between node $i$ in layer $l$ and node $i$ in layer $l \pm 1$), the static graph representation as an adjacency matrix can be expanded to a dynamic graph representation as an adjacency tensor \cite{mucha2010community,bassett2013robust}, providing important mathematical advantages to common statistical challenges present in these data.

The tools of temporal networks are particularly useful in modeling plasticity and learning with the aim of predicting recovery \cite{rienkensmeyer2016computational}. In initial efforts, temporal networks have been used to reveal common patterns of network reconfiguration that occur as healthy adult individuals learn a new motor-visual skill over the course of days to weeks \cite{bassett2011dynamic,bassett2013task,mantzaris2013dynamic,bassett2014cross,bassett2015learning}. Interestingly, individuals that displayed greater network flexibility, particularly in areas of the brain critical for cognitive control \cite{bassett2015learning} learned more quickly than individuals with less flexible brains \cite{bassett2011dynamic}.  While these studies initially applied temporal network techniques to motor skill learning with the aim of informing rehabilitation after stroke \cite{heitger2012motor}, there are many open questions about how sensitive these techniques might be to neural or cognitive plasticity underlying other types of learning \cite{mattar2016network,karuza2016local}, or to other dynamic processes that are of important to neuroengineering in other clinical contexts.

Beyond temporal networks, one can extend the multilayer network construct to represent relationships across different imaging modalities \cite{muldoon2016network}: for example, calcium transients and local field potentials, or structural MRI and EEG, or diffusion imaging and functional MRI \cite{nicosia2014spontaneous}. Alternatively, one could think of letting each layer represent a different frequency band \cite{brookes2016multi,domenico2016mapping} or a different patient in a clinical cohort. Indeed, the potential applications of these multilayer representations across neuroimaging contexts is surprisingly broad, and future efforts will likely include a careful assessment of their utility in uncovering conserved and variable properties of networked neural systems.

\subsection*{Statistical tools, frameworks, and null models}

An important burgeoning area of work lies in building, testing, and validating appropriate statistical methods and models for network inference. Because networks are not simple mathematical objects, the tools required to capture and compare them extend beyond what traditional statistics offers \cite{kolaczyk2009statistical}. Many efforts have focused on developing sophisticated permutation-based methods for network comparison \cite{simpson2013permutation,winkler2015multi}, and some have extended these methods to assess differences in network \emph{functions} (rather than univariate statistics) \cite{ginestet2011statistical,bassett2012altered,betzel2016modular}, for example by building on tools developed in the field of functional data analysis \cite{ramsay2005functional,ramsay2002applied,ramsay2006functional}. In addition to comparing networks, one often wishes to understand whether the network structure or dynamics that one observes in empirical data is expected or unexpected. Answering these questions depends on the development of appropriate static and dynamic network null models (see \cite{betzel2016modular,papadopoulos2016evolution,bassett2015extraction} and \cite{bassett2013robust}, respectively). Statistical considerations also extend to estimating the connectivity itself \cite{lindquist2014evaluating,lepage2013inferring}, assessing its significance \cite{ginestet2011statistical}, and measuring its relationship to behavior or symptomatology \cite{shehzad2014multivariate}. Finally, a nascent area of inquiry lies in building statistical models of networks  \cite{simpson2011exponential,simpson2012exponential,klimm2014resolving} in order to understand their generative principles \cite{betzel2016generative,vertes2012simple,vertes2014generative,pavlovic2014stochastic}.

\subsection*{Algebraic topology}

\begin{figure}[h]
\begin{center}
\centerline{\includegraphics[width=0.45\textwidth]{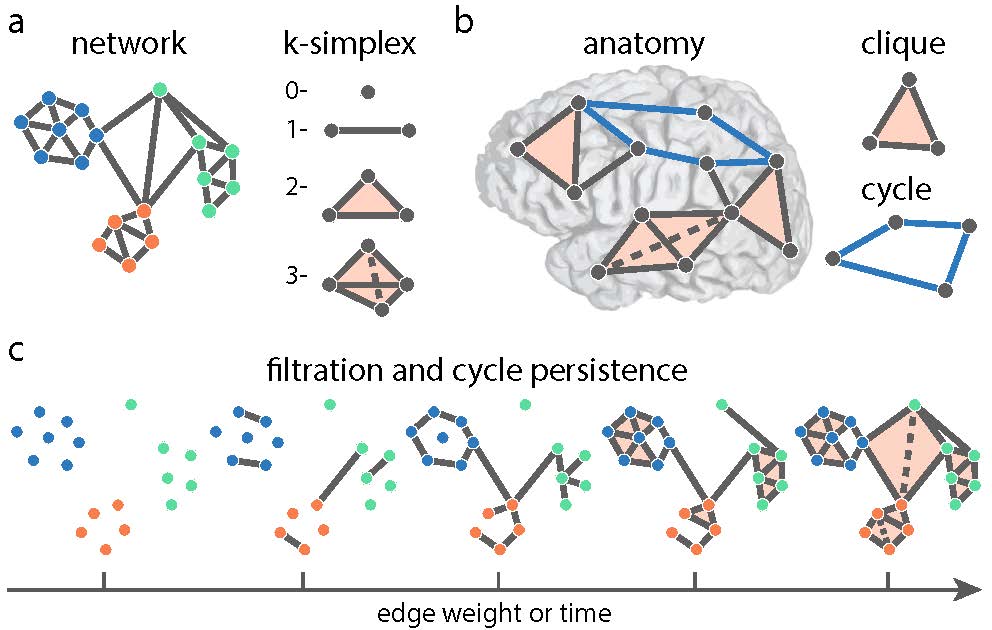}}
\caption{\textbf{Tools for Higher-Order Interactions from Algebraic Topology.} \textbf{(a)} The human connectome is a complex network architecture that contains both dyadic and higher-order interactions. Graph representations of the human connectome only encode dyadic relationships, and leave the higher-order interactions unaccounted for. A natural way in which to encode higher-order interactions is in the language of algebraic topology, which defines building blocks called simplices \cite{giusti2016twos}: a 0-simplex is a node, a 1-simplex is an edge between two nodes, a 2-simplex is a filled triangle, etc. \textbf{(b)} These building blocks enable the desciption of two distinct structural motifs that are thought to play very different roles in neural computations \cite{curto2016what}: (i) cliques, which are all-to-all connected subgraphs, are thought to facilitate integrated codes and computations, and (ii) cycles or cavities, which are collection of $n$-simplices arranged to have an empty geometric boundary, are thought to facilitate segregated codes and computations. \textbf{(c)} Additional tools available to the investigator include filtrations and persistent homology. Filtrations represent weighted simplicial complexes as a series of unweighted simplicial complexes, and can be used to study networks that change over time, or that display hierarchical structure across edge weights.  Filtrations allow one to follow cycles from one complex to another and quantify how long they live (via the number of complexes in which they are consecutively present). Because this is a study of the persistence of a cycle, it is referred to as the persistent homology of the weighted simplicial complex. \label{fig4_algebraic_topology}}
\end{center}
\end{figure}

While extremely powerful, network science is largely built on the tools of graph theory, which inherently treat the dyad (a single connection between two nodes) as the fundamental unit of interest. Recent evidence, however, points to the fact that sensor networks, technological networks, and even neural networks display higher-order interactions that simply cannot be reduced to pairwise relationships \cite{ganmor2011architecture,ganmor2011sparse}. To address this growing realization, we can turn to recent advances in applied algebraic topology \cite{ghrist2014elementary}, which reframes the problem of relational data in terms of \emph{simplices} or collections -- rather than pairs -- of vertices \cite{giusti2016twos} (Fig.~\ref{fig4_algebraic_topology}). This added sensitivity enables algebro-topological tools to offer mechanisms for neural coding \cite{giusti2015clique,curto2016what}, distinguish disparate classes of graph models \cite{sizemore2016classification}, and separate healthy from clinical populations \cite{kim2014morphological}. The framework also offers useful tools to consider the evolution of simplices over time drawing on the notion of a \emph{filtration}, and tools to identify and track hollow cavities in networks -- structures that are otherwise invisible to common graph metrics \cite{giusti2016twos}. We anticipate that the next few years will see an increasing interest in better understanding the role of these higher order interactions in healthy cognition \emph{versus} disease, and their sensitivity as biomarkers for tracking effects of training and rehabilitation.

\subsection*{Network control theory}

\begin{figure}[h]
\begin{center}
\centerline{\includegraphics[width=0.45\textwidth]{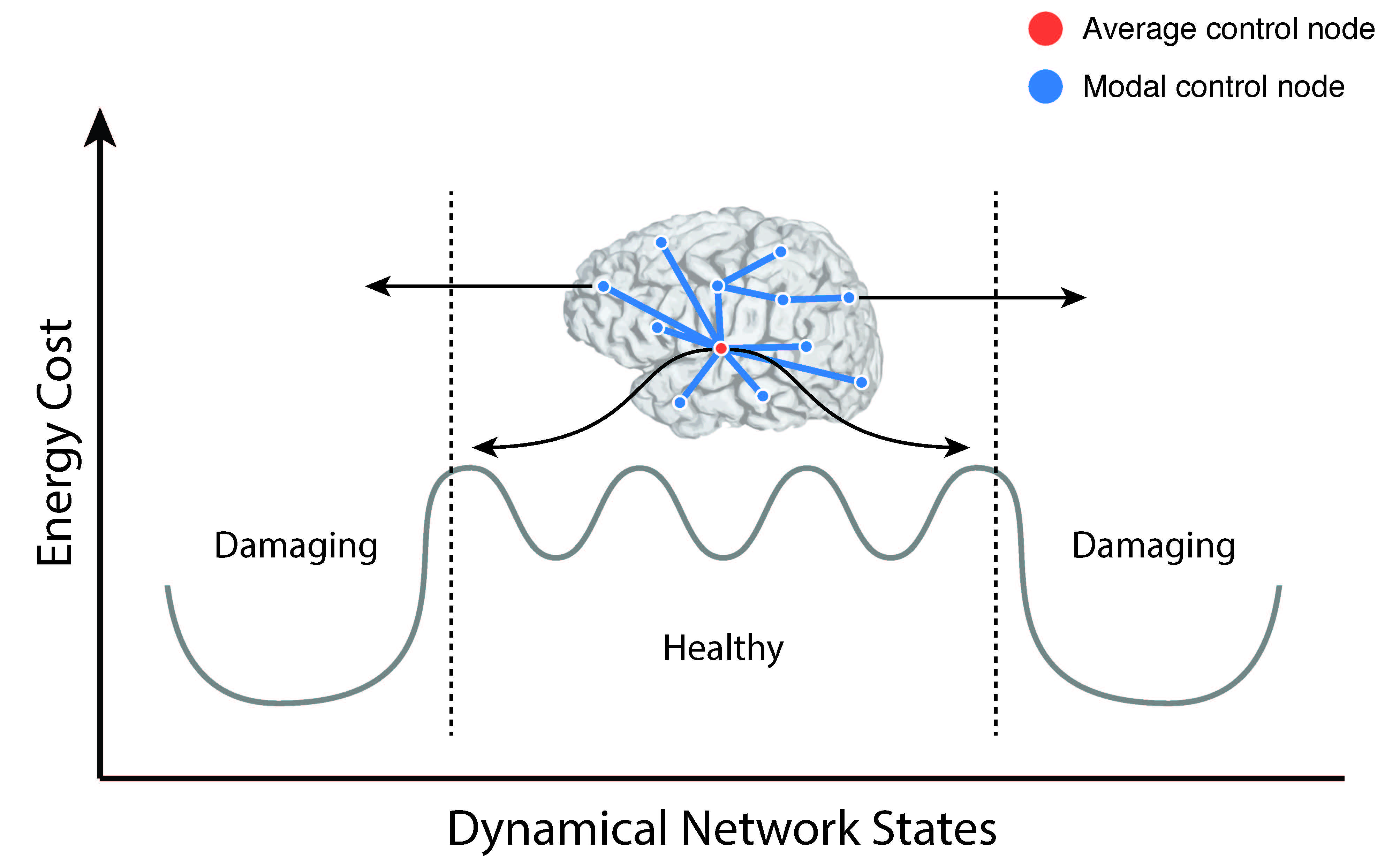}}
\caption{\textbf{Brain network regulation and control can help navigate dynamical states.} To accomplish behavioral and cognitive goals, brain networks internally navigate a complex space of dynamical states. Putative brain states may be situated in various peaks and troughs of an energy landscape -- requiring the brain to expend metabolic energy to move from the current state to the next state. Within the space of possible dynamical states, there are easily accessible states and harder-to-reach states; in some cases, the accessible states are healthy while in other cases they may contribute to dysfunction, and similarly for the harder-to-reach states. Two commonly observed control strategies used by brain networks are average control and modal control. In average control, highly central nodes navigate the brain towards easy-to-reach states. In contrast, modal control nodes tend to be isolated brain regions that navigate the brain towards hard-to-reach states that may require additional energy expenditure \cite{gu2015controllability}. As a self-regulation mechanism for preventing transitions towards damaging states, the brain may employ cooperative and antagonistic, push-pull, strategies \cite{khambhati2016virtual}. In such a framework, the propensity for the brain to transition towards a damaging state might be competitively limited by opposing modal and average controllers whose goal would be to pull the brain towards less damaging states.} \label{fig3_control_theory}
\end{center}
\end{figure}

A final exciting frontier that we will mention -- which bridges both computational science and systems engineering -- is the development and application of explicitly \emph{network} control theory \cite{liu2011controllability,pasqualetti2014controllability} to neural systems (Fig.~\ref{fig3_control_theory}). Indeed, neural control engineering \cite{schiff2011neural} is slowly evolving into neuro-network control engineering, as the control problems become tuned to the underlying graph architecture of the dynamical processes \cite{motter2015network}. In general, these applications take one of two forms: either seeking to understand how neural systems control themselves, or how one can exogeneously control a neural system, steering it away from pathological dynamics and towards healthy dynamics.

In the first case, we seek to understand how neural systems control themselves. To address this question, we can write down a model of brain dynamics where the current brain state depends on (i) the previous brain state, (ii) the wiring pattern that structurally connects network nodes (brain regions), and (iii) the control input. Assuming this is a linear, time-invariant, noise-free, and discrete-time model, we can infer which brain regions are predisposed to affect the system, and in what ways. Early efforts along these lines revealed that regions in the brain's executive system are well-poised to push the brain into difficult-to-reach states, far away on an energy landscape \cite{gu2015controllability}. Moreover, the brain's densely interconnected rich-club is poised to form the ground state of the system, being the least energetically costly target state \cite{betzel2016optimally}. Interestingly, these control principles of the brain, built on the organization of white matter tracts, are significantly altered in individuals who have experienced traumatic brain injury \cite{gu2016optimal}, suggesting their utility in clinical applications. However, it is also important to be cautious; these predictions are based on \emph{linear} network control while the brain is a nonlinear dynamical system, and therefore interpretations should be validated in additional studies \cite{cornelius2013realistic}. For example, demonstrating that individual differences in cognitive control function are correlated with individual differences in network control statistics will be an important first step \cite{medaglia2016cognitive}, as will demonstrating that these statistics change over developmental time scales in which cognitive control emerges in children \cite{tang2016structural}.  Moreover, exploring the applicability of nonlinear control strategies, including linearization of nonlinear systems, will be an important avenue of inquiry for future work \cite{cornelius2013realistic}.

The second context in which network control theory offers a powerful toolset for neuroengineers is in addressing the question of how to exogeneously control a neural system and accurately predict the outcome on neurophysiological dynamics -- and, by extention, cognition and behavior. Indeed, how to target, tune, and optimize stimulation interventions is one of the most pressing challenges in the treatment of Parkinson's disease and epilepsy, to name a couple \cite{johnson2013neuromodulation}. More broadly, this question directly impacts the targeting of optogentic stimulation in animals \cite{ching2013control} and the use of invasive and non-invasive stimulation in humans \cite{muldoon2016stimulation} (e.g., deep brain, grid, transcranial magnetic, transcranial direct current, and transcranial alternating current stimulation). As a case study, consider medically refractory epilepsy, where network techniques can be used to identify seizure onset zones \cite{hao2014computing,khambhati2015dynamic,burns2014network} and where network control theory can be used to detect drug-resistant seizures \cite{santaniello2011quickest}, inform the development of a distributed control algorithm to quiet seizures using grid stimulation \cite{ching2012distributed}, and identify areas of the network to target during resective surgery \cite{khambhati2016virtual}.

\section*{Towards Clinical Translation}

\begin{figure}[h]
\begin{center}
\centerline{\includegraphics[width=0.45\textwidth]{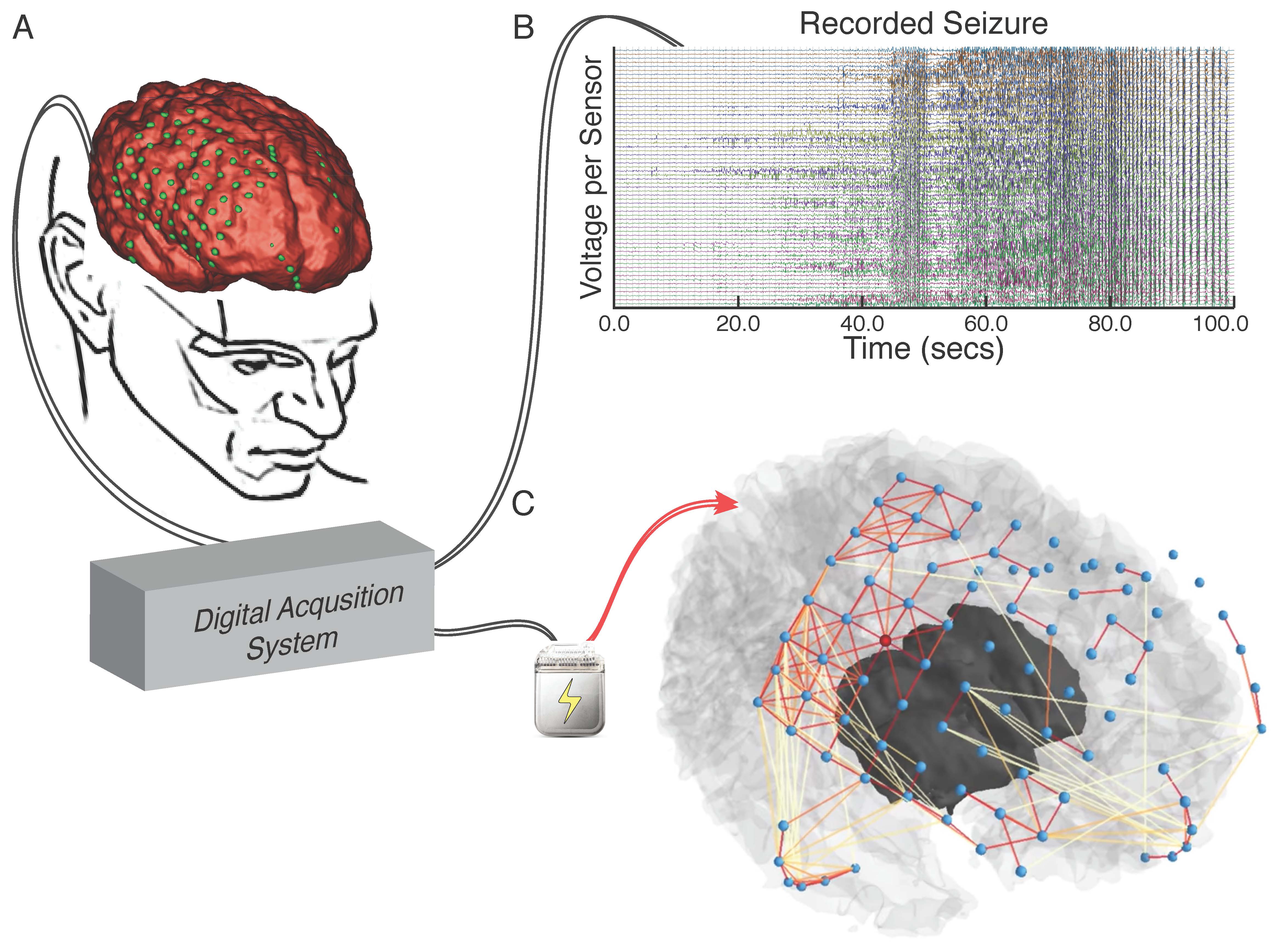}}
\caption{\textbf{Clinical translation of network neuroscience tools.} Network neuroscience offers a natural framework for improving tools to diagnose and treat brain network disorders. (\textbf{A}) For drug-resistant epilepsy patients, invasive monitoring of brain activity to localize brain tissue where seizures originate and plan resective surgery is challenging, because the neural processes generating seizures are poorly understood. Epileptic brain signals, electrical fields produced by the firing of neuron populations, are sensed by electrodes that rest on the surface of the brain, beneath the dura, and are recorded by a digital acquisition system. A three-dimensional reconstruction of a patient's brain (red) with electrodes co-localized (green) to anatomical features is shown here. (\textbf{B}) Recorded brain signals are studied by clinical practitioners to characterize spatial and temporal behavior of the patient's seizure activity. In the plot, each line represents time-varying voltage fluctuation from each electrode sensor. (\textbf{C}) Inferred functional connections from a single time-slice during the patient's recorded seizure demonstrates rich relationships in neural dynamics between brain regions and are not visually evident from \textbf{B} (blue circles are nodes, red links are strong connections, yellow links are weak connections). Functional connectivity patterns demonstrate strong interactions around brain regions in which seizures begin and weak projections to brain regions where seizures spread. Objective tools in network neuroscience can usher in an era of personalized algorithms capable of mapping epileptic network architecture from neural signals and pinpointing implantable, neurostimulation devices to specific brain regions for intervention \cite{muldoon2016stimulation,khambhati2015dynamic,khambhati2016virtual}.} \label{fig5_lastsection}
\end{center}
\end{figure}

Together, these exciting computational frontiers have the potential to directly inform clinical practice. Indeed, several of the main translational challenges of neuroengineering are ripe for the incorporation of network data. These opportunities begin at the earliest stages of clinical diagnosis and monitoring, where variation between individuals -- and even indeed variation within a single individual -- stymie progress in tuning medication, stimulation, brain-machine interfaces, neuroprosthetics, and physical or cognitive-behavioral therapy to offer individuals a better quality of life (Fig.~\ref{fig5_lastsection}). Concerted efforts in mapping relational architectures in neural and behavioral data in the form of graphs and networks will be critical to obtaining a more holistic understanding of mental health, as well as greater insights into optimizing interventions. Such mappings could occur in the traditional sense using empirical measurements performed in research or hospital settings; but perhaps the most tantalizing possibilities currently being discussed include the use of digital data from smart phones to accurately phenotype individuals, and the health of their nervous system, with the goal of better guiding intervention strategies for the clinically unwell \cite{onnela2016harnessing,torous2016new}.

\section*{Extensions Beyond Neuroengineering}

Before concluding, it is important to point out that the mathematical methods and conceptual frameworks that we have been discusing in this review are more generally applicable beyond the specific realm of brain connectivity. From genes \cite{beagan2016local} to the musculo-skeletal system, from central to peripheral nervous systems \cite{chen2016neural}, and from injured neural tissue in brains causing cognitive deficits to neural tissue in muscles causing pain \cite{zhang2016stretch}, network science offers an approach that spurns reductionism in favor of wholistic maps and models of complex interconnected systems. Indeed, future work may benefit from considering the nervous system as embedded or embodied in the broader context, as only one part of an interconnected web of networks supporting human life \cite{baldassano2016topological,steinway2015inference}.

\section*{Conclusion and Future Outlook}

In this review, we have sought to introduce an exciting and emerging frontier in neuroengineering: a network science of brain connectivity. In addition to outlining the mathematical underpinnings of the field, we have briefly described some marked initial successes in which the tools of network neuroscience have been brought to bear on neural mapping and connectivity estimation, diagnosis and monitoring, and rehabilitation and treatment. However, we are also careful to describe common pitfalls and associated limitations, in an effort to offer a balanced guide in incorporating these techniques into one's own research practices. We took the liberty to speculate in the later sections about some important frontiers that we believe will become increasingly critical to the questions posed by neuroengineering in the near future, both from a computational point of view and from a view towards clinical impact. In closing, we underscore yet again that the strength and novelty of network neuroscience lies in its brazen grasp on the full complexities of relational data, facilitating transformative approaches to understanding, fixing, and building brains.
~\\
~\\
\emph{Acknowledgements.} We thank Chad Giusti, Jason Kim, and Matthew Hemphill for helpful comments on an earlier version of this manuscript. DSB would also like to acknowledge support  from the John D. and Catherine T. MacArthur Foundation, the Alfred P. Sloan Foundation, the  Army Research Laboratory and the Army Research Office through contract numbers W911NF-10-2-0022 and W911NF-14-1-0679,the National Institute of Mental Health (2-R01-DC-009209-11), the National Institute of Child Health and Human Development  (1R01HD086888-01), the Office of Naval Research, and the National Science Foundation (BCS-1441502, BCS-1430087, BCS-1631550, and CAREER PHY-1554488). The content is solely the responsibility of the authors and does not necessarily represent the official views of any of the funding agencies.

\bibliography{bibfile}

\end{document}